\documentclass[aps,preprint]{revtex4}%
\usepackage[dvips]{color}
\usepackage{amsfonts}
\usepackage{amsmath}
\usepackage{amssymb}
\usepackage{graphicx}%
\providecommand{\U}[1]{\protect\rule{.1in}{.1in}}
\newtheorem{theorem}{Theorem}
\newtheorem{acknowledgement}[theorem]{Acknowledgement}

\newcommand{\ket}[1]{\left|{#1}\right\rangle}

\begin{document}
\preprint{ }
\title[quantum information storage]{Spectral hole burning for stopping light}
\author{R. Lauro, T. Chaneli\`ere, J.- L. Le Gou\"{e}t}
\affiliation{Laboratoire Aim\'{e} Cotton, CNRS UPR3321, Univ. Paris Sud, bat\^{\i}ment 505,
campus universitaire, 91405 Orsay, France}

\begin{abstract}
We propose a novel protocol for storage and retrieval of photon wave packets in a $\Lambda$-type atomic medium. This protocol derives from spectral hole burning and takes advantages of the specific properties of solid state systems at low temperature, such as rare earth ion doped crystals. The signal pulse is tuned to the center of the hole that has been burnt previously within the inhomogeneously broadened absorption band. The group velocity is strongly reduced, being proportional to the hole width. This way the optically carried information and energy is carried over to the off-resonance optical dipoles. Storage and retrieval are performed by conversion to and from ground state Raman coherence by using brief $\pi$-pulses. The protocol exhibits some resemblance with the well known electromagnetically induced transparency process. It also presents distinctive features such as the absence of coupling beam. In this paper we detail the various steps of the protocol, summarize the critical parameters and theoretically examine the recovery efficiency.         

\end{abstract}
\volumeyear{year}
\volumenumber{number}
\issuenumber{number}
\eid{identifier}
\date[Date text]{date}
\received[Received text]{date}

\revised[Revised text]{date}

\accepted[Accepted text]{date}

\published[Published text]{date}

\startpage{1}
\endpage{ }
\maketitle
\section{Introduction}
Future long distance quantum information networks are expected to involve quantum repeaters. To this end, intense investigation is going on all over the world to interface a quantum state of light with an atomic ensemble. In protocols based on electromagnetically induced transparency (EIT), the input signal is adiabatically converted into an atomic ground state coherence, without excitation of optical coherences. This class of protocols has been investigated intensively both theoretically \cite{fleisch2,lukin,fleisch} and experimentally, leading to storage and retrieval demonstration of both discrete \cite{chan,eisa} and continuous \cite{appel,honda} quantum variables. As it bypasses the optical coherence excitation, the EIT based protocol is well adapted to atomic vapors, with strong transitions and short optical dipoles lifetime.

Rare earth ion doped crystals (REIC) also offer long ground state coherence lifetime, which makes them attractive for quantum storage. However, in contrast with the atomic vapors considered in EIT experiments, they also exhibit long optical coherence times. In addition, the active ions are strictly motionless, the hyperfine sublevel relaxation time may reach tens of hours, and the optical transition frequency is distributed over a very broad inhomogeneous width.

Because of these specific features EIT-based protocols may not be the most appropriate approach to quantum storage in this type of medium. Actually, different protocols have been considered recently, based on rephasing processes\cite{nils,kraus,afze,hetet,alex,ried}. In the present paper, we propose a new adiabatic protocol, taking advantage of the specific REIC properties. 

It is well known that a light pulse slows down when it propagates through a transparency window (TW) in an absorbing medium. Deeper and narrower the transparency window, slower the light pulse. The group velocity $v$ is of the order $\Delta_0 / \alpha_0$, where $\Delta_0$ and $\alpha_0$ respectively stand for the TW width and the linear absorption coefficient outside the TW. Provided the input signal spectrum is narrower than the transparency window, the pulse undergoes little energy loss and shape distortion. 

In EIT, the TW is opened by a strong classical coupling field, interacting with the atoms simultaneously with the input signal. As the input signal is slowed down to group velocity $v<<c$, the pulse length and energy are reduced by a factor $v/c<<1$. The pulse energy is carried over to the coupling field, and the input signal quantum state is mapped into the ground state Raman coherence. This coherence behaves as a spin wave that adiabatically follows \cite{grisch} the signal pulse, at the same velocity $v$. By switching off the coupling beam one closes the TW. This simultaneously destroys the slowly propagating pulse and freezes the spin wave that already contains most of the signal information \cite{mats}. To retrieve the stored information, one turns on the coupling beam. Interacting with the coupling beam, the frozen spin wave restores the signal pulse. Conversely, the reviving light pulse drives the spin wave to the medium output where the initial energy is finally returned to the light signal by the coupling beam.   

In REIC, relying on the homogeneous width narrowness, the hyperfine relaxation slowness, and the absence of translational motion, one can prepare a narrow TW, by frequency selective optical pumping to a ground state shelving sublevel. This is nothing but the well known spectral hole-burning process. With an hyperfine level lifetime ranging from a few seconds to several hours, the preparation step can be completed long before the arrival of the input signal. After the preparation step, an input signal that is sent through the spectral hole undergoes the same slowing down as in EIT \cite{shakh}. In contrast with EIT, the energy and the signal state are both carried over to the optical coherence of the off-resonance atoms. The transient optical dipoles are created along the path of the signal field. Their natural lifetime is much longer than the pulse duration, but they vanish adiabatically as the light pulse passes by. To ``stop'' the light and store the information, we propose to convert those optical dipoles into Raman coherences, while the slowed down pulse is still confined within the absorbing medium. This can be performed by an auxiliary $\pi$-pulse, much shorter than the inverse TW width. Another $\pi$-pulse shall restore the optical dipoles and release the stored information.

The paper is arranged as follows. We describe slowing down and storage in section II. Section III is devoted to the signal recovery. The latter section represents the core of the paper. We detail the revival of the signal from the restored optical coherence. Specifically we determine the rising time of the retrieved signal, and its temporal profile. We finally compute the recovery efficiency, taking account of the size of the absorbing medium and of the signal spectral width.

\section{Slowing down and storage}
\subsection{The slowing down step}
\begin{figure}
\includegraphics[width=8cm]{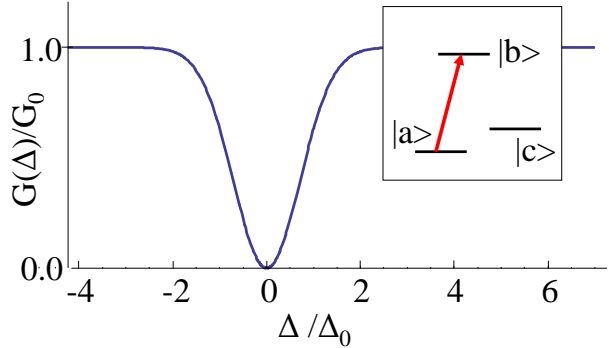}
\caption{(color online) inhomogeneous distribution $G(\Delta)/G_0$ with a $\Delta_0$-wide spectral hole. The inset represents the $\Lambda$-shape three-level system. The input signal pulse excites transition a-b. }
\label{fig_distribution}
\end{figure}

We consider an ensemble of motion-less three-level atoms in a $\Lambda$ configuration. The excited state $\ket b$ is connected by optical transitions to the ground substates $\ket a$ and $\ket c$. The light pulse to be stored interacts with transition $\ket a-\ket b$ (see Fig. \ref{fig_distribution}).  The inhomogeneous broadening of the optical line is much larger than the pulse width. A hole is burnt in the absorption profile over an interval $\Delta_0$ around frequency $\omega_0$. This is achieved by pumping the corresponding atoms  to some additional long lifetime shelving level. The homogeneous linewith $\gamma_{ab}$ is assumed to be much smaller than $\Delta_0$. All these assumptions are consistent with the properties of some rare earth ion doped crystals at low temperature. The absorbing atoms are distributed as a function of the transition frequency $\omega_{ab}$. Since $\gamma_{ab}<<\Delta_0$ various shapes can be given to the burnt hole. 

Slow light with persistent hole burning was already discussed thoroughly \cite{shakh} and we just summarize the main features relevant to the proposed storage protocol. The incoming field $E(z,t)$, centered at frequency $\omega_0$, propagates along direction $z$ and is expressed in terms of the slowly varying envelope $\mathcal{A}(z,t)$ as: 
\begin{equation}
E(z,t)=\frac{1}{2}\left(\mathcal{A}(z,t)\mathrm{e}^{i(\omega_0 t-k z)}+c.c.\right )%
\end{equation}
The pulse propagation is described by the linearized wave equation:
\begin{equation}\label{wave_equation01}
\partial_z \mathcal{A}(z,t)+\frac{1}{c}\partial_t \mathcal{A}(z,t)=-i\frac{k}{\epsilon_0}\mathcal{P}(z,t)
\end{equation}
The polarization density is expressed as the sum $\mathcal{P}(z,t)=\mu_{ab}\int d\Delta G(\Delta)\sigma_{ab}(\Delta;z,t)$ of the off-diagonal density matrix element 
\begin{equation}\label{rho}
\sigma_{ab}(\Delta;z,t)=-i\frac{\mu_{ab}}{2\hbar}\int_0^\infty d\tau \mathcal{A}(z,t-\tau)\mathrm{e}^{(i\Delta-\gamma_{ab})\tau}
\end{equation}
where $\mu_{ab}$ stands for the transition dipole moment. The sum is performed over the distribution $G(\Delta)$ of the atomic transition frequency $\omega_{ab}=\omega_0+\Delta$. The distribution $G(\Delta)$, initially uniform over a broad interval, has been depressed by hole burning over a $\Delta_0$-wide spectral window centered at $\Delta=0$, where $G(0)=0$.      

In the weak field limit, the equation is most conveniently solved in spectral domain in terms of linear susceptibility. Time-to-frequency Fourier transform of Eq. \ref{wave_equation01} leads to:
\begin{equation}\label{wave_equation02}
\partial_z \tilde{\mathcal{A}}(z,\Omega)+i\frac{\Omega}{c}\tilde{\mathcal{A}}(z,\Omega)=-i\frac{k}{\epsilon_0}\tilde{\mathcal{P}}(z,\Omega)
\end{equation} 
where:
\begin{equation}
\tilde{\mathcal{P}}(z,\Omega)=\frac{1}{2}\epsilon_0\chi(\omega_0+\Omega)\tilde{\mathcal{A}}(z,\Omega)
\end{equation}
and 
\begin{equation}\label{susceptibility}
\chi(\omega_0+\Omega)=\frac{\mu_{ab}^2}{\hbar\epsilon_0}\int d\Delta \frac{G(\Delta+\Omega)}{\Delta+i\gamma_{ab}} 
\end{equation}
The absorption coefficient can be expressed in terms of the imaginary part $\chi"(\omega)$ of the susceptibility as $\alpha(\omega)=-\chi"(\omega)/k$. At large spectral distance from the burnt hole, this quantity reduces to:
\begin{equation}\label{coeff_absorption}
\alpha_0=\pi G_0 \frac{k\mu_{ab}^2}{\hbar\epsilon_0},
\end{equation} 
where $G_0$ stands for the uniform inhomogeneous distribution away from the hole. 

To get more tractable expressions in the burnt hole region, we expand the susceptibility to second order in $\Omega$. Then Eq. \ref{wave_equation02} reduces to:
\begin{equation}\label{wave_equation03}
\partial_z \tilde{\mathcal{A}}(z,\Omega)+\left[i\frac{\Omega}{v}+\frac{1}{2}\alpha(\omega_0+\Omega)\right]\tilde{\mathcal{A}}(z,\Omega)=0
\end{equation} 
where the absorption coefficient and the inverse group velocity are respectively expressed as:
\begin{equation}\label{absorption}
\alpha(\omega_0+\Omega)=\alpha_0\left(\int d\Delta g(\Delta)\mathcal{L}(\Delta)+\frac{\Omega^2}{2}\int d\Delta \frac{\partial^2g(\Delta)}{\partial \Delta^2}\mathcal{L}(\Delta)\right)
\end{equation}   
\begin{equation}\label{group_v}
\frac{1}{v}=\frac{1}{c}+\frac{\alpha_0}{2\pi}\int d\Delta g(\Delta)\frac{\Delta^2}{(\Delta^2+\gamma_{ab}^2)^2}
\end{equation}
where $g(\Delta)=G(\Delta)/G_0$ denotes the normalized inhomogeneous distribution and $\mathcal{L}(\Delta)=\pi^{-1}\gamma_{ab}/(\Delta^2+\gamma_{ab}^2)$ represents the Lorentzian homogeneous line profile.
  
If the incoming pulse spectrum is much narrower than $\Delta_0$, absorption is well described by the $\Omega$-independent coefficient $\alpha(\omega_0)$. Then, by Fourier transforming Eq. \ref{wave_equation03} solution back to time domain one obtains:   
\begin{equation}\label{solution_wave1}
\mathcal{A}(z,t)=\mathcal{A}(0,t-z/v)\mathrm{e}^{-\frac{1}{2}\alpha(\omega_0)z}
\end{equation}
This is a travelling wave propagating without distortion at velocity $v$ through the $L$-long slab. According to Eqs. \ref{absorption} and \ref{group_v}, $\alpha(\omega_0)$ and $1/v$ are respectively of order $\alpha_0 \gamma_{ab}/\Delta_0$ and $\alpha_0/\Delta_0$. The wave is not attenuated provided $\alpha_0 L\gamma_{ab}/\Delta_0<<1$. With respect to free space, the wave envelope is spatially compressed by factor $v/c$ when light travels inside the atomic medium, the pulse duration $T$ remaining unchanged. 

The entire pulse may be confined inside the medium if the group delay $L/v$ is larger than $T$, which also reads $\Delta_0T<<\alpha_0 L $. This condition must be consistent with the width of the burnt hole, since a lower limit to the pulse spectral width is given by $1/T$. To account for the finite size of the transparency window we use the second order term in Eq. \ref{absorption}. At the hole center, $\partial_{\Delta}^2g(\Delta)\approx\Delta_0^{-2}$. Then, with $\gamma_{ab}<<\Delta_0$, the transmission factor at spectral distance $\Omega$ from the hole center reads as $\mathrm{exp}(-\frac{1}{2}\alpha_0 L \Omega^2/\Delta_0^2)$. Hence, the pulse spectral width is much smaller than the transmission window width if $\Delta_0T>>\sqrt{\alpha_0 L}$. The two conditions can be combined into:   
\begin{equation}
\sqrt{\alpha_0 L}>>1,
\end{equation}
showing that the optical density is the only critical parameter. Not surprisingly, a similar condition has been derived for the EIT process\cite{fleisch2,lukin}.

To check the validity of the susceptibility second order expansion, let us consider a burnt hole with a gaussian profile. Then, provided $\gamma_{ab}<<\Delta_0$, the susceptibility reads as:
\begin{equation}\label{chi_gauss}
\chi(\omega_0+\Omega)=\frac{\alpha_0}{k}\left[-i\left(1-\mathrm{e}^{-\frac{\Omega^2}{\Delta_0^2}}\right)+\frac{2}{\sqrt{\pi}}\mathrm{F}(\frac{\Omega}{\Delta_0})\right]
\end{equation}     
where $\mathrm{F}(x)$ represents the Dawson integral. To second order in $\Omega$, Eq. \ref{chi_gauss} can be expanded as:
\begin{equation}\label{chi_expansion}
\chi(\omega_0+\Omega)=\frac{\alpha_0}{k}\left(\frac{2}{\sqrt{\pi}}\frac{\Omega}{\Delta_0}-i\frac{\Omega^2}{\Delta_0^2}\right)
\end{equation}
\begin{figure}
\includegraphics[width=8cm]{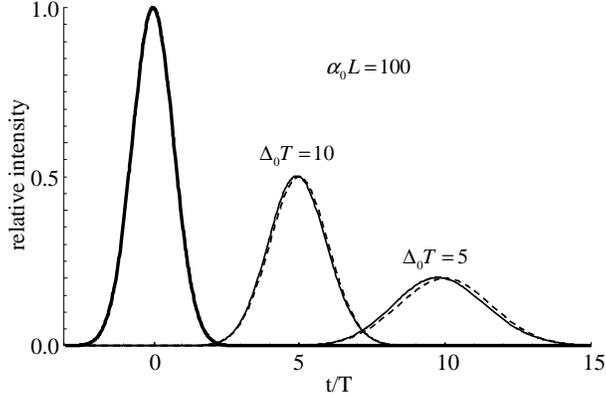}
\caption{ propagation of a gaussian pulse through a gaussian spectral hole. The bold, solid and dashed lines respectively represent the initial pulse shape, the exactly calculated transmitted pulse shape and a second order susceptibility approximation of the transmitted pulse profile. Opacity is $\alpha_0L=100$ and the displayed profiles correspond to $\Delta_0T=5$ and $\Delta_0T=10$.}
\label{fig_propag}
\end{figure}
As illustrated in Fig. \ref{fig_propag}, the propagation of a gaussian pulse, as computed to second order approximation with the help of Eqs. \ref{wave_equation03} and \ref{chi_expansion}, keeps very close to the exact solution derived from Eqs \ref{wave_equation02} and \ref{chi_gauss}.
\subsection{The storage step}
Being spatially compressed in the medium, the signal pulse looses most of its energy. If the medium is comprised of the active atoms alone, the dielectric host being ignored, the index of refraction is unity at $\omega_0$ and the field amplitude is conserved at the input and output sides of the slab. The spatial density of energy is also conserved. By summing this density over the compressed envelope, one can easily show that the optically carried energy is reduced by the factor $v/c$. The same phenomenon occurs in EIT. In this process energy flows from the incoming pulse to the coupling beam at the input side of the medium and is restored by the coupling beam at the output side. In the hole burning process the situation is different since there is no coupling beam. Then the atoms located outside the burnt hole serve as a temporary energy tank. As discussed a long time ago by D. Grischkowsky \cite{grisch}, this transient off-resonance excitation can be analyzed in terms of adiabatic following. 

With the condition $\lim_{t \rightarrow -\infty}\mathcal{A}(z,t)=0$, repeated integration by parts of Eq. \ref{rho} leads to:
\begin{equation}\label{adiabatic_following}
\sigma_{ab}(\Delta;z,t) = -\frac{1}{2} \frac{\mu_{ab}}{\hbar}\sum_{n=0}^{\infty} \left(\frac{-i}{ \Delta +i \gamma_{ab}}\right)^{n+1} \frac{\partial^{n}}{\partial t^{n}} \mathcal{A}(z,t)
\end{equation}
To lowest order in $\mu_{ab} \mathcal{A}(z,t)/\hbar/\Delta$, off-resonance atom coordinates in the Bloch sphere equator read as:
\begin{equation}
u(\Delta;z,t) = -\frac{\mu_{ab}}{\hbar}\frac{\mathcal{A}(z,t)}{ \Delta } ,v(\Delta;z,t) = -\frac{\mu_{ab}}{\hbar}\frac{\partial_t\mathcal{A}(z,t)}{ \Delta^2 }  
\end{equation} 
where $\sigma_{ab}(\Delta;z,t) =\left[u(\Delta;z,t)+iv(\Delta;z,t)\right]/2$.  Therefore, despite the weakness of the driving field, the off resonance optical dipoles with different detunings do not dephase from each other, as long as the field is present. On the contrary, in the absence of the field, the position in the Bloch sphere equator would rotate at angular velocity $\Delta$.

\begin{figure}
\includegraphics[width=8cm]{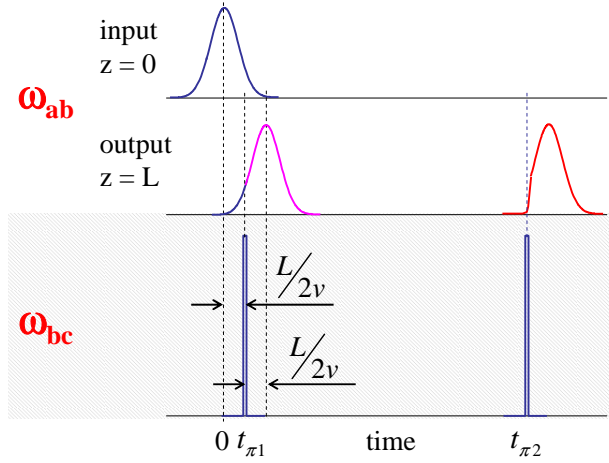}
\caption{(color online) time diagram for storage and retrieval. Incoming and restored signal temporal profiles are represented in the input and ouput sides of the memory medium. The incoming pulse is tuned to the burnt hole center, interacting with the atoms on transition $\ket a -\ket b$. The storage $\pi$-pulse is applied at time $t_{\pi1}=L/(2v)$ on transition $\ket b -\ket c$, when the incoming pulse is halfway from the slab ends. Retrieval is accomplished by a second $\pi$-pulse at time $t_{\pi2}$. The restored profile is represented in red. In the absence of storage pulse, the original pulse emerges from the slab with the delay $L/v$ (magenta line). The same color code is used in Fig. \ref{fig_restored}.}
\label{fig_time_diagram}
\end{figure}

The time diagram of the proposed storage and retrieval protocol is illustrated in Fig. \ref{fig_time_diagram}. By shining a brief $\pi$-pulse on transition b-c one can convert the optical dipoles into Raman coherence. Illumination by the $\pi$-pulse must take place at a moment $t_{\pi1}$ when the signal pulse is confined in the medium. A pulse of duration $\tau_c$ successfully converts dipoles up to $1/\tau_c$ from resonance. This way one decouples the atoms from the signal pulse. Since the compressed signal pulse carries a small part of the initial energy, the information is really frozen inside the medium. Deprived of the co-propagating dipole wave, the signal pulse recovers the velocity of light in vacuum and exits the medium. Indeed it takes a time of order $1/\Delta_0$, much larger than $L/c$, to rebuild the macroscopic polarization needed to slow down a light pulse. The pulse accelerates to $c$ as a whole, which means its duration is reduced by the factor $v/c$ and its spectral width is conversely magnified by the factor $c/v$.      
     
\section{Recovery of the stored signal}
\subsection{Expression of the retrieved signal}
In order to restore the signal field one applies a second brief $\pi$-pulse at time $t_{\pi2}$ (see Fig. \ref{fig_time_diagram}). Calculating the restored signal represents the core of this paper. As discussed in section II, the atomic dipoles are locked to the incoming pulse in the first stage of the process. When they are restored by the second $\pi$-pulse, they come to life in a field-free medium. Since no field is yet able to lock them, they rapidly dephase from each other. At the same time, they give rise to a radiative response that will be able, ultimately, to keep them in phase. One may wonder whether dephasing may affect the restored signal amplitude.

At the moment, let us assume that the conversion operated by the two $\pi$-pulses is perfect. The recovered polarisation can be expressed as:
\begin{equation}
\mathcal{P}_1(z,t)=-i\frac{\mu_{ab}^2}{2\hbar}Y(t-t_{\pi2})\int d\Delta G(\Delta)\int_0^\infty d\tau  \mathcal{A}_{in}(z,t_{\pi1}-\tau)\mathrm{e}^{(i\Delta-\gamma_{ab})(\tau+t-t_{\pi2})}
\end{equation}
where $\mathcal{A}_{in}(z,t)$ represents the initial field amplitude and $Y(t)$ denotes the unit step function. The polarisation also includes the atomic reaction to the restored field $\mathcal{A}_{out}(z,t)$:
\begin{equation}
\mathcal{P}_2(z,t)=-i\frac{\mu_{ab}^2}{2\hbar}\int d\Delta G(\Delta)\int_0^\infty d\tau \mathcal{A}_{out}(z,t-\tau)\mathrm{e}^{(i\Delta-\gamma_{ab})\tau}
\end{equation}
The wave equation for $\mathcal{A}_{out}(z,t)$ reads as:
\begin{equation}\label{wave2}
\partial_z \mathcal{A}_{out}(z,t)+\frac{1}{c}\partial_t \mathcal{A}_{out}(z,t)=-i\frac{k}{\epsilon_0}\left[\mathcal{P}_1(z,t)+\mathcal{P}_2(z,t)\right]
\end{equation}
In the same way as in Eqs \ref{wave_equation02}-\ref{susceptibility}, Fourier conjugation leads to:
\begin{equation}\label{wave_equation_recovery}
\partial_z \tilde{\mathcal{A}}_{out}(z,\Omega)+i\left(\frac{\Omega}{c}+\frac{1}{2}\epsilon_0\chi(\omega_0+\Omega)\right)\tilde{\mathcal{A}}_{out}(z,\Omega)=-i\frac{k}{\epsilon_0}\tilde{\mathcal{P}_1}(z,\Omega)
\end{equation} 
Solving the equation and transforming back to the time domain, one obtains:
\begin{equation}\label{restored_signal_1}
\mathcal{A}_{out}(z,t)=-\frac{\alpha_0}{2\pi}\int d\Delta g(\Delta)\int_0^\infty d\tau\int_0^\infty d\tau' \mathrm{e}^{(i\Delta-\gamma_{ab})(\tau+\tau')}\mathcal{R}(t_{\pi1}-\tau,t-t_{\pi2}-\tau',z)
\end{equation}
where
\begin{equation}\label{restored_signal_2}
\mathcal{R}(t',t,z)=\int_0^z dz'\mathcal{A}_{in}(z',t')\int\frac{d\Omega}{2\pi}\mathrm{e}^{i\Omega t-i\left(\frac{\Omega}{c}+\frac{1}{2}k\chi(\omega_0+\Omega)\right)(z-z')}
\end{equation}
\subsection{Restored signal rising time}
Some time is needed for the signal to revive. First we verify that it starts from zero at $t=t_{\pi2}$. We  specify the hole profile assuming $g(\Delta)= 1-\mathrm{exp}(-\Delta^2/\Delta_0^2)$. As illustrated in Fig. \ref{fig_propag}, the susceptibility can be reduced to its second order expansion in $\Omega$.  Then, as shown in the Appendix, the restored signal can be written as:
\begin{equation}\label{restored_signal_main}
\begin{split}
\mathcal{A}_{out}(z,t)=-\frac{\alpha_0v}{2\pi}\int d\Delta g(\Delta)\int_0^\infty d\tau\int_{\max(0,t-t_{\pi2}-z/v)}^{t-t_{\pi2}} d\tau' \mathrm{e}^{(i\Delta-\gamma_{ab})(\tau+\tau')}\\\sum_{n=0}^\infty\frac{1}{n!}\left[\beta\frac{v}{\Delta_0}\right]^n  \partial_{z'}^{2n}\left[\mathcal{A}_{in}(z',t_{\pi 1}-\tau)(z-z')^n\right]_{z'=z-v\theta}
\end{split}
\end{equation}
where $\theta=t-t_{\pi2}-\tau'$. This expression vanishes at $t=t_{\pi2}$ since the interval of integration over $\tau'$ reduces to $0$. Next we determine the signal rising time. 

\begin{figure}
\includegraphics[width=8cm]{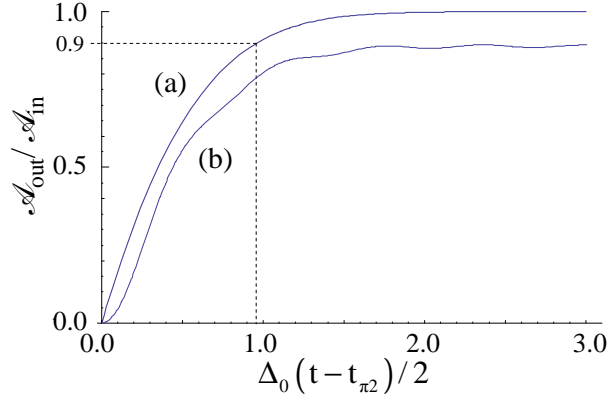}
\caption{Evolution of the retrieved signal amplitude after illumination by the second $\pi$-pulse at $t_{\pi2}$. (a) infinite bandwith operation. (b) finite bandwith operation: $\Delta_1=5\Delta_0$. The wavy line appearance reflects a sharp cut-off in the frequency domain}
\label{fig_recovery}
\end{figure}

The restored signal is observed at the slab output, at $z=L$. The $n=0$ term dominates in Eq. \ref{restored_signal_main} provided:
\begin{equation}
\mathcal{A}_{in}(z',t_{\pi 1}-\tau)>>\beta\frac{v}{\Delta_0}\partial_{z'}^2\left[\mathcal{A}_{in}(z',t_{\pi 1}-\tau)(L-z')\right]_{z'=L-v\theta}     
\end{equation}
Let $\mathcal{A}_{in}(z,t)$ be a Fourier transform limited pulse of duration $T$. Given that the pulse spectrum is assumed to be narrower than the burnt hole, and that $\partial_{z}^2\mathcal{A}_{in}(z,t)\approx \mathcal{A}_{in}(z,t)/(vT)^2$, the condition reduces to:
\begin{equation}\label{condition}
\Delta_0\min\left(t-t_{\pi2},L/v\right)<<(\Delta_0T)^2    
\end{equation}   
Within the time interval when this condition is satisfied, the restored field is given by:
\begin{equation}\label{simple_field}
\mathcal{A}_{out}(L,t)=-\frac{\alpha_0v}{2\pi}\int d\Delta g(\Delta)\int_0^\infty d\tau\int_{\max(0, t-t_{\pi2}-L/v)}^{t-t_{\pi2}} d\tau'\mathrm{e}^{(i\Delta-\gamma_{ab})(\tau+\tau')}\mathcal{A}_{in}(L-v\theta,t_{\pi 1}-\tau)
\end{equation}
The sum over $\tau'$ indicates that only slices at distance $<v(t-t_{\pi2})$ contribute to the restored signal. Let us first focus on the time region $t<t_{\pi2}+L/v$. Then the lower boundary of the integral over $\tau'$ is $0$. Significant contributions only come from $\tau$ and $\tau'$ values smaller than $\Delta_0^{-1}$. Given that $\Delta_0T>>1$, we can consider that $\mathcal{A}_{in}(L-v\theta,t_{\pi 1}-\tau)$ does not vary on this timescale. Then, taking $\mathcal{A}_{in}(L-v\theta,t_{\pi 1}-\tau)$ out of the integrals at $\tau=\tau'=0$ one obtains:
\begin{equation}\label{revival}
\mathcal{A}_{out}(L,t)=\mathcal{A}_{in}(L-v(t-t_{\pi2}),t_{\pi 1})\kappa\left[(t-t_{\pi2})\Delta_0/2\right]
\end{equation}
where
\begin{eqnarray}
\kappa(x)&=&-\frac{\alpha_0v}{2\pi}\int d\Delta g(\Delta)\int_0^\infty d\tau\int_0^{2x/\Delta_0} d\tau'\mathrm{e}^{(i\Delta-\gamma_{ab})(\tau+\tau')}\\
&=&\left(1-\frac{v}{c}\right)\left(1-\mathrm{e}^{-x^2}+x\sqrt{\pi}\left[1-\mathrm{erf}(x)\right]\right)
\end{eqnarray}
As illustrated in Fig.\ref{fig_recovery}, the field recovery is 90$\%$ completed at time $t=t_{\pi2}+2/\Delta_0$.

Thus far we have assumed that storage and retrieval are accomplished with an infinite bandwidth, all the atoms contributing to the process, whatever their distance $\Delta$ from the burnt hole. The atoms contribute with a weight $1/\Delta^2$ to the restored signal, as can be seen by considering the $(t-t_{\pi2})\Delta_0>>1$ limit of Eq. \ref{revival}:
\begin{equation}
\mathcal{A}_{out}(L,t)=\mathcal{A}_{in}(L-v(t-t_{\pi2}),t_{\pi 1})\frac{\alpha_0v}{2\pi}\int d\Delta g(\Delta)\frac{1}{(\Delta+i\gamma_{ab})^2}
\end{equation}  
Therefore, finite bandwidth operation should reduce the recovery efficiency by a factor close to $1-\Delta_0/\Delta_1$, where $2\Delta_1$ denotes the process bandwidth. With the same gaussian hole as above, this factor equals $\mathrm{erf}(y)-\frac{1}{\sqrt{\pi}}[1-\mathrm{exp}(-y^2)]/y$ where $y=\Delta_1/\Delta_0$. The corresponding time evolution of the restored field is shown in Fig. \ref{fig_recovery} for $\Delta_1=5\Delta_0$. The wavy line appearance results from the sharp cut-off in the frequency domain. 
\subsection{Established signal}
\begin{figure}
\includegraphics[width=8cm]{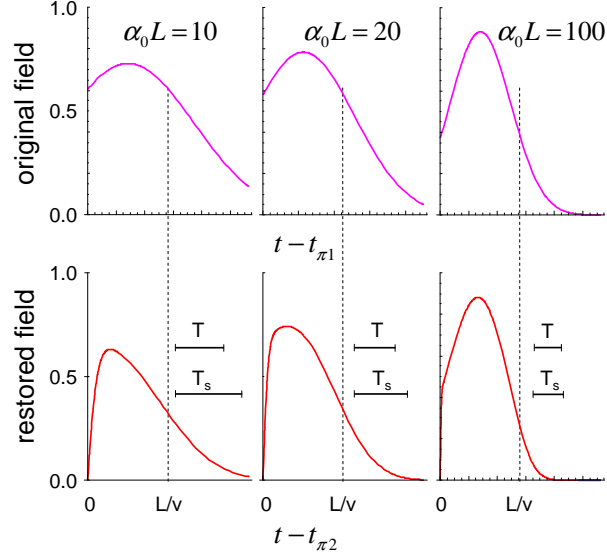}
\caption{ (color online) original (magenta lines) and restored (red lines) field as a function of $t-t_{\pi1}$ and $t-t_{\pi2}$ respectively for different values of $\alpha_0L$. The original pulse is gaussian and reaches $1$, its maximum value, at the origin of time at the slab input. Storage is performed at $t_{\pi1}=L/(2v)$. One adjusts the pulse duration to the opacity of the medium by setting: $\Delta_0T=0.6(\alpha_0L)^{3/4}$. The original and stretched pulse durations $T$ and $T_s$ are indicated for each $\alpha_0L$ value. The group delay $L/v$ determines the temporal memory depth of the slab. }
\label{fig_restored}
\end{figure}

Although the series expansion in Eq. \ref{restored_signal_main} was convenient for studying the signal birth, this expression is difficult to handle when Eq. \ref{condition} is no longer satisfied. A different approach is then needed. Once the signal is established, the quantity $\mathcal{R}(t_{\pi1}-\tau,t-t_{\pi2}-\tau',z)$ evolves on a timescale much longer than $\Delta_0^{-1}$ and can be taken out of the integral at $\tau=\tau'=0$ in Eq. \ref{restored_signal_1}. The restored signal amplitude reduces to:
\begin{equation}\label{restored_signal_3}
\mathcal{A}_{out}(L,t)=\left(1-\frac{v}{c}\right)\mathcal{R}(t_{\pi1},t-t_{\pi2},L)
\end{equation}  
To go further we have to specify the incoming pulse profile. We proceed with the gaussian field: 
\begin{equation}
\mathcal{A}_{in}(0,t)=\mathrm{e}^{-\frac{t^2}{2T^2}}
\end{equation} 
According to Eq. \ref{wave_equation03}, propagation from the input side to $z'$ changes the field amplitude into:
\begin{equation}\label{transmitted_field}
\mathcal{A}_{in}(z',t)=\frac{\Delta_0T}{\sqrt{\Delta_0^2T^2+\alpha_0z'}}\mathrm{e}^{-\frac{\Delta_0^2\left(t-\frac{z'}{v}\right)^2}{2(\Delta_0^2T^2+\alpha_0z')}}
\end{equation} 
Substituting Eq. \ref{transmitted_field} into Eq. \ref{restored_signal_2} and using the susceptibility second order expansion one obtains:
\begin{equation}
\mathcal{R}(\frac{x}{\Delta_0},\frac{y}{\Delta_0},\frac{\rho v}{\Delta_0})=\int_0^{\frac{\rho v}{\Delta_0}} du\frac{\Delta_0T}{\sqrt{2\pi a(\rho-u)}\sqrt{\Delta_0^2T^2+au}}\mathrm{e}^{-\frac{(x-u)^2}{2(\Delta_0^2T^2+au)}-\frac{(\rho-u-y)^2}{2a(\rho-u)}}
\end{equation}
where $a=\sqrt{\pi}$. To optimize recovery, we assume that storage occurs when the incoming pulse sits halfway from the slab ends, which means $t_{\pi1}=L/(2v)$. In order to satisfy the double condition $\sqrt{\alpha_0L}<<\Delta_0T<<\alpha_0L$, which is needed to confine the pulse within the slab both spatially and spectrally, we set $\Delta_0T=b(\alpha_0L)^{3/4}$, where $b$ is close to unity. In Fig. \ref{fig_restored} we display the time evolution of both the original and the restored signals for various values of the optical density. The upper boxes represent the restored field as a function of $t-t_{\pi2}$ at the slab output. The lower boxes represent the initial field as a function of $t-t_{\pi1}$, in the absence of storage and recovery pulses. The computation is performed with $b=0.6$. In the figure we indicate the size of $T$ for each value of $\alpha_0L$. Propagation through the burnt hole affects the pulse spectral wings, stretching the temporal profile. The size of the corresponding elongated pulse width, denoted $T_s$ and given by $T_s =T\left[1+\alpha_0L/(\Delta_0T)^2\right]1/2$, is also indicated in the figure. As expected, as $\alpha_0L$ is increased, a larger part of the incoming signal is comprised within the temporal storage window $L/v$. In other words, as $\alpha_0L$ is made bigger, a smaller part of the initial field leaves the slab before the storage time $t_{\pi1}$, or has not yet entered the slab at time $t_{\pi1}$. With growing $\alpha_0L$, the restored field is less distorted with respect to the incoming signal. Finally, the restored field drops rapidly at time longer than $L/v$, since the finite size of the slab truncates the trailing edge of the incoming pulse. 
   
It should be noticed that Eq. \ref{simple_field} is not valid only in the leading edge of the restored signal. According to Eq. \ref{condition}, this simple form of the retrieved field is valid for any value of $t-t_{\pi2}$ provided $\Delta_0L/v<<(\Delta_0T)^2$. Since, according to Eq. \ref{simple_field}, the signal mainly comes from depth $v(t-t_{\pi2})$, it rapidly vanishes when $v(t-t_{\pi2})$ exceeds the slab dimension $L$.    
\subsection{Recovery efficiency}
The recovery efficiency is simply defined as the ratio of the incoming and restored energies:
\begin{equation}
\eta=\frac{\int_{t_{\pi2}}^{\infty}dt\left|\mathcal{A}_{out}(L,t)\right|^2 }{\int_{-\infty}^{\infty}dt\left|\mathcal{A}_{in}(0,t)\right|^2}
\end{equation} 
This quantity is displayed in Fig. \ref{fig_recovery} as a function of $\sqrt{\alpha_0L}$. The computation is performed with the same parameters as in Fig. \ref{fig_restored}. The expected recovery efficiency behavior is similar to what is observed in EIT.  
\begin{figure}
\includegraphics[width=8cm]{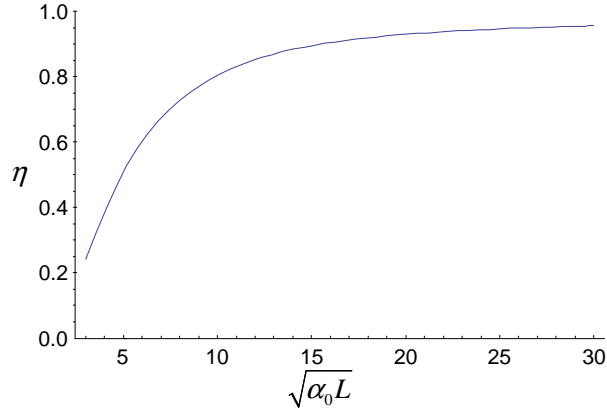}
\caption{ recovery efficiency as a function of $\sqrt{\alpha_0L}$. The original pulse is gaussian. Storage is performed at $t_{\pi1}=L/(2v)$. One adjusts the pulse duration to the opacity of the medium by setting: $\Delta_0T=0.6(\alpha_0L)^{3/4}$. }
\label{fig_efficiency}
\end{figure}
 
The question of the control field dynamic adjustment in EIT experiments was raised recently \cite{gorsh}. It was shown that the temporal variation of the control field at writing and retrieval could be optimized to improve the recovery efficiency. According to these works, one can map an arbitrary input pulse on the optimal spin wave, unique for each opacity value, by appropriate temporal shaping of the control field during the writing step. Conversely, the stored information can be retrieved in any desired temporal shape with adequate profiling of the readout control field. In the present situation we apparently lose this degree of freedom since the burnt hole cannot be reshaped during the storage process. However, the $z$ spatial coordinate plays a role similar to the time coordinate. A $z$-dependent spectral hole profile might act on the input pulse in a similar way as the time dependent control field in EIT. Such a $z$-dependent profile can be achieved by transverse illumination of the slab \cite{shakh,tsch}.       
\section{conclusion}
Relying on the specific properties of REIC, we have proposed a new optical storage protocol that is based on persistent spectral hole burning. As in EIT, the input pulse propagates with negligible
absorption and is adiabatically coupled to a collective atomic state coherence. In contrast with EIT, inhomogeneous broadening plays a crucial role and information is initially carried over to off-resonance optical dipoles. Storage relies on the conversion of optical dipoles into Raman coherence and back with the help of brief auxilliary $\pi$-pulses. Requirements are similar to those of EIT, and the expected efficiency is also similar. With respect to EIT, the main practical feature is the absence of coupling beam. The restored signal is temporally separated from the read out $\pi$-pulse. Another interesting feature is the spectral separation of the atoms that carry the information and those that are involved in the hole burning process. We have derived a semi-classical theory but the extension to quantum storage would be straightforward.  
\begin{acknowledgement}
The authors are pleased to acknowledge stimulating discussions with Paul R. Berman.
\end{acknowledgement}        
\appendix
\renewcommand\theequation{A.\arabic{equation}}
\section{Restored field calculation}  

This calculation aims at using the properties of the Dirac peak and its derivatives. With an inhomogeneous distribution given by $g(\Delta)= 1-\mathrm{exp}(-\Delta^2/\Delta_0^2)$, the sum over $\Omega$ in Eq. \ref{restored_signal_2} can be expressed in terms of the susceptibility second order expansion as:
\begin{equation}
\int\frac{d\Omega}{2\pi}\mathrm{e}^{i\Omega t-i\left(\frac{\Omega}{c}+\frac{1}{2}k\chi(\omega_0+\Omega)\right)(z-z')}=\int\frac{d\Omega}{2\pi}\mathrm{e}^{i\Omega(t-\frac{z-z'}{v})-\beta\frac{\Omega^2}{\Delta_0}\frac{z-z'}{v}}
\end{equation} 
where $\beta=\frac{\sqrt{\pi}}{2}\left(1-\frac{v}{c}\right)$. To extract the Dirac peak one expands the exponential in the following way:
\begin{equation}
\int\frac{d\Omega}{2\pi}\mathrm{e}^{i\Omega(t-\frac{z-z'}{v})-\beta\frac{\Omega^2}{\Delta_0}\frac{z-z'}{v}}=\sum_{n=0}^\infty\frac{1}{n!}\left[\frac{\beta}{\Delta_0}\frac{z-z'}{v}\right]^n\int\frac{d\Omega}{2\pi}(-)^n\Omega^{2n}\mathrm{e}^{i\Omega(t-\frac{z-z'}{v})}
\end{equation}
With the help of the Dirac peak derivative definition:
\begin{equation}
\int\frac{d\Omega}{2\pi}(i\Omega)^{2n}\mathrm{e}^{i\Omega t}=\frac{d^{2n}}{dt^{2n}}\int\frac{d\Omega}{2\pi}\mathrm{e}^{i\Omega t}=\delta^{(2n)}(t)
\end{equation}
and of the property:
\begin{equation}
\int dt f(t)\delta^{(2n)}(t)=\frac{d^{2n}}{dt^{2n}}f(0)
\end{equation}
one can perform the sum over $z'$ in Eq. \ref{restored_signal_2} as:
\begin{equation}\label{Dirac_series}
\int_0^z\frac{dz'}{v}\mathcal{A}_{in}(z',t')\left(\frac{z-z'}{v}\right)^n\delta^{(2n)}\left(t-\frac{z-z'}{v}\right)=v^n\Pi_{0,z/v}(t)  \partial_{z'}^{2n}\left[\mathcal{A}_{in}(z',t')(z-z')^n\right]_{z'=z-vt}
\end{equation}
where $\Pi_{x,y}(t)=1$ when $t\in[x,y]$ and vanishes outside. Finally, substituting Eq. \ref{Dirac_series} into Eq. \ref{restored_signal_2} one obtains:
\begin{equation}
\begin{split}
\mathcal{A}_{out}(z,t)=-\frac{\alpha_0v}{2\pi}\int d\Delta g(\Delta)\int_0^\infty d\tau\int_{\max(0,t-t_{\pi2}-z/v)}^{t-t_{\pi2}} d\tau' \mathrm{e}^{(i\Delta-\gamma_{ab})(\tau+\tau')}\\\sum_{n=0}^\infty\frac{1}{n!}\left[\beta\frac{v}{\Delta_0}\right]^n  \partial_{z'}^{2n}\left[\mathcal{A}_{in}(z',t_{\pi 1}-\tau)(z-z')^n\right]_{z'=z-v\theta}
\end{split}
\end{equation} 
where $\theta==t-t_{\pi2}-\tau'$.


\begin{thebibliography}{99}                                                                                               %
\bibitem {fleisch2}M. Fleischhauer and M.D. Lukin, Phys. Rev. Lett.
\textbf{84}, 5094 (2000); Phys. Rev. A \textbf{65}, 022314 (2002).

\bibitem {lukin}M. D. Lukin, Rev. Mod. Phys. \textbf{75}, 457 (2003), and references therein. 

\bibitem {fleisch}Michael Fleischhauer, Atac Imamoglu, and Jonathan P.
Marangos, Rev. Mod. Phys. \textbf{77}, 633 (2005), and references therein.

\bibitem {chan}T. Chaneli\`{e}re, D.N. Matsukevich, S.D. Jenkins, S.-Y. Lan,
T.A.B. Kennedy, and A. Kuzmich, Nature \textbf{438}, 833 (2005).

\bibitem {eisa}M.D. Eisaman, A. Andr\'e, F. Massou, M. Fleischhauer, A.S.
Zibrov, M.D. Lukin, Nature \textbf{438}, 837 (2005)

\bibitem {honda}K. Honda, D. Akamatsu, M. Arikawa, Y. Yokoi, K. Akiba, S.
Nagatsuka, T. Tanimura, A. Furusawa and M. Kozuma, Phys. Rev. Lett.
\textbf{100}, 093601 (2008)

\bibitem {appel}J. Appel, E. Figueroa, D. Korystov, M. Lobino and A.I. Lvovsky, Phys.
Rev. Lett. \textbf{100}, 093602 (2008)

\bibitem {nils}M. Nilsson and S. Kr\"{o}ll, Opt. Commun. {\textbf 247} 393(2005). 

\bibitem {kraus}B. Kraus, W. Tittel, N. Gisin, M. Nilsson, S. Kr\"{o}ll, and
J. I. Cirac, Phys. Rev. A \textbf{73}, 020302(R) (2006)

\bibitem {afze}Mikael Afzelius, Christoph Simon, Hugues de Riedmatten, Nicolas
Gisin, arXiv:0805.4164

\bibitem {hetet}G. H\'{e}tet, J. J. Longdell, A. L. Alexander, P. K. Lam, and
M. J. Sellars, Phys. Rev. Lett. \textbf{100}, 023601 (2008); J. J. Longdell,
G. H\'{e}tet, P. K. Lam, and M. J. Sellars, Phys. Rev. A \textbf{78}, 032337 (2008).

\bibitem {alex}A. L. Alexander, J. J. Longdell, M. J. Sellars, and N. B.
Manson, Phys. Rev. Lett. \textbf{96}, 043602 (2006).

\bibitem {ried}Hugues de Riedmatten, Mikael Afzelius, Matthias Staudt,
Christoph Simon, Nicolas Gisin, arXiv:0810.0630

\bibitem {grisch}D. Grischkowsky, Phys. Rev. A {\textbf 7} 2096(1973). 

\bibitem {mats}A. B. Matsko, Y. V. Rostovtsev, O. Kocharovskaya, A. S. Zibrov, and M. O. Scully, Phys. Rev. A {\textbf 64} 043809 (2001); A. S. Zibrov, A. B. Matsko, O. Kocharovskaya, Y. V. Rostovtsev, G. R. Welch, and M. O. Scully, Phys. Rev. Lett. {\textbf 88} (2002) 103601 

\bibitem {shakh}R. N. Shakhmuratov, A. Rebane, P. M\'egret, and J. Odeurs, Phys. Rev. A {\textbf 71} 053811 (2005).  

\bibitem {gorsh}A. V. Gorshkov, A. Andr\'e, M. Fleischhauer, A. S. Sorensen, and M. D. Lukin, Phys. Rev. Lett. {\textbf 98} 123601 (2007); A. V. Gorshkov, A. Andr\'e, M. D. Lukin and A. S. Sorensen, Phys. Rev. A {\textbf 76} 033805 (2007); I. Novikova, N. B. Phillips, and A. V. Gorshkov, Phys. Rev. A {\textbf 78} 021802(R) (2008). 

\bibitem {tsch}M. Tschanz, A. Rebane, and U. P. Wild, Opt. Eng. (Bellingham) {\textbf 34} 1936 (1995); M. Tschanz, A. Rebane, D. Reiss, and U. P. Wild, Mol. Cryst. Liq. Cryst. Sci. Technol., Sect. A {\textbf 283} 43 (1996).



\end{thebibliography}
\end{document}